\documentclass[aps,showpacs,preprint,superscriptaddress]{revtex4}
\usepackage{graphicx}
\begin{document}
\bibliographystyle{revtex}
\title{A microscopic study of the proton-neutron symmetry and phonon structure of the low-lying
states in $^{92}$Zr}
\vspace{0.5cm}
\author{N. Lo Iudice}
\email{loiudice@na.infn.it}
\affiliation{Dipartimento di Scienze Fisiche, Universit\'a di Napoli 
"Federico II" \\and 
Istituto Nazionale di Fisica Nucleare,\\
Monte S Angelo, Via Cinthia I-80126 Napoli, Italy}
\author{Ch. Stoyanov}
\email{stoyanov@inrne.bas.bg}
\affiliation{ Institute for Nuclear Research and Nuclear Energy,
1784 Sofia, Bulgaria}
\date{\today}
\begin{abstract}
We studied in a microscopic multiphonon approach the
proton-neutron symmetry and phonon structure of some low-lying
states recently discovered in $^{92}$Zr. We confirm the breaking
of F-spin symmetry, but argue that the breaking mechanism is more
complex than the one suggested in the original shell model
analysis of the data. We found other new intriguing features of
the spectrum, like a pronounced multiphonon fragmentation of the
states and a tentative evidence of a
three-phonon mixed symmetry state.
\end{abstract}
\maketitle

\section{Introduction}

The experiments on $^{94}$Mo \cite{Pietr99,Pietr00,Frans01,Frans03},
$^{96}$Ru \cite{Pietr01,Klein02} and other nuclei \cite{Gade02}
have provided a conclusive evidence in favor of the existence of
F-spin mixed symmetry states in nearly spherical nuclei, thereby
confirming a major prediction of the proton-neutron interacting
boson model (IBM2)\cite{Ari77}. The many levels and transition
probabilities produced by the experiments fit extraordinarily well
in the IBM2 scheme according to the F-spin quantum number and the
number of quadrupole bosons. Indeed, states with the same F-spin
symmetry differing by one-phonon were found to be strongly coupled
by isoscalar E2 transitions, while $F=F_{max}$ symmetric and
$F=F_{max}- 1$ mixed symmetry states with the same number of
phonons were connected through strong M1 transitions, a feature
that qualifies the mixed-symmetry states as scissors-like
excitations \cite{LoPa78,Bohle84} built on quadrupole vibrational
states.

The phonon structure as well as the proton-neutron (p-n) symmetry
were confirmed by a microscopic calculation using a
multiphonon basis, whose phonons were generated in random-phase
approximation (RPA) \cite{LoSto00,LoSto02}. This approach, known
as quasiparticle-phonon model (QPM) \cite{Sol92}, accounted with
good accuracy for the energies, transition probabilities and
selection rules. A good  description was also given by a truncated
shell model (SM) calculation \cite{Lise00}.

In the attempt of establishing how far the F-spin symmetry holds
valid, the same experimental group performed a photon scattering
experiment \cite{Werner02} on $^{92}$Zr, which is supposed to be
at the border line of the F-spin symmetry domain, and analyzed the
data in the framework of the SM adopted in Ref. \cite{Lise00} for
studying the $^{94}$Mo. The most meaningful outcome of their
analysis was that the two lowest $2^{+}$ states are one-phonon
excitations, at variance with the other nuclei of the region where
the second $2^{+}_{2}$ is a two-phonon  state. More
remarkably, the first $2^{+}_{1}$ is basically a pure neutron
excitation, while the second $2^{+}_{2}$ is purely isovector.

This latter point, implying a severe breaking of F-spin,  gave us
the main motivation for the present QPM study. This approach, in
fact, using explicitly a RPA phonon basis and taking into account
a very large number of configurations excluded in SM, provides a
more stringent microscopic test for the the validity of the F-spin
symmetry in $^{92}$Zr.

\section{Calculation and results}

We used the same Woods-Saxon potential and the same
separable two-body Hamiltonian adopted in
\cite{LoSto00,LoSto02}. We used also the same single particle basis, 
which encompasses all bound states from the bottom of the well up to
the quasi-bound states embedded into the continuum. 
Following the same strategy, we fit the
strength $\kappa_{2}$ of the quadrupole-quadrupole (Q-Q)
interaction on the energy and E2 decay strength of the $2^{+}_{1}$
and the coupling constant $G_{2}$ of the quadrupole pairing on the
overall properties of the low-lying $2^{+}$ isovector state. As we 
shall see, this
came out to be $G_{2}\simeq \kappa_{2}$, slightly larger than in
the case of $^{94}$Mo ($G_{2}\simeq 0.8 \kappa_{2}$). The other
Hamiltonian parameters remained unchanged. Because of the large model 
space, we used effective
charges very close to the bare values, namely 
$e_{p}=1.1$ for protons and $e_{n} = 0.1$ for neutrons.  
We also used the spin-gyromagnetic quenching
factor $g_{s} = 0.7$.

\subsection{A very brief outline of the procedure}
Following the QPM procedure,  we transformed  the
Hamiltonian into the phonon form
\begin{eqnarray}
H_{QPM}  =  \sum_{i\mu } \omega_{i\lambda } Q_{i \lambda  \mu }^{\dagger}
Q_{i \lambda  \mu } + H_{vq}\label{Hphon},
\end{eqnarray}
where the first term is the unperturbed phonon Hamiltonian and  $ H_{vq}$
is a phonon-coupling piece whose exact expression can
be found in Ref. \cite{Sol92}. Both terms are
expressed in terms of the RPA phonon operators
\begin{equation}
Q^{\dagger}_{i\lambda \mu }\,=\,\frac{1}{2}
 \sum_{jj'} \left \{ \psi_{jj'}^{i \lambda }
[\alpha^{\dagger}_{j} \alpha^{\dagger}_{j'}]_{\lambda \mu}
-(-1)^{\lambda - \mu} \varphi_{jj'}^{i \lambda }
[\alpha_{j'}\alpha_{j}]_{\lambda -\mu} \right \}
\label{ph}
\end{equation}
of multipolarity $\lambda \mu$ and
energy $\omega_{i \lambda }$,
where $\alpha^{\dagger}_{jm} (\alpha_{jm})$  are quasiparticle
operators obtained from the corresponding particle operators
through a Bogoliubov transformation.
The phonon operators fulfil the normalization conditions
\begin{equation}
\frac{1}{2}
\sum_{j  j'} \left[ \psi_{jj'}^{i \lambda }
\psi_{jj'}^{i' \lambda }-
 \varphi_{jj'}^{i \lambda }\varphi_{jj'}^{i' \lambda' }\right]
 =\delta_{ii'} \delta_{\lambda \lambda'}.
\label{phnorm}
\end{equation}
It is worth to point out that, among the RPA phonons, 
only few are collective, composed of a coherent linear 
combination of two-quasiparticle configurations. 
The Boson Hamiltonian is then accordingly diagonalized
in a space spanned by RPA multiphonon states. As in Ref.
\cite{LoSto00,LoSto02}, we included up to three-phonon states. 

\subsection{RPA analysis}
The preliminary condition for QPM being able to test the symmetry
properties and phonon composition of the low-lying states as
described in IBM2, is the occurrence at low energy of at least two
RPA quadrupole phonons, which we denote  $[2^{+}_{1}]_{RPA}$
and $[2^{+}_{2}]_{RPA}$. 
In the specific case of $^{92}$Zr, even in QPM   
the two lowest lying states are to be one-phonon states,
the first dominated by  $[2^{+}_{1}]_{RPA}$, the second by
$[2^{+}_{2}]_{RPA}$.
To test the proton-neutron symmetry of these RPA states we compute the ratios
\begin{eqnarray}
R_{i=1,2}(IV/IS) = \frac{{\cal M}^{(n)}_{i} -
{\cal M}^{(p)}_{i}}{{\cal M}^{(n)}_{i} +
{\cal M}^{(p)}_{i}},\,\,\,\,\,\,\,\,\,\,\,\,\,\,\,  
R^{2}_{i}(IV/IS) =  \vert\, R_{i}(IV/IS)\, \vert^{2},
\end{eqnarray}
where
\begin{equation}
{\cal M}^{(n/p)}_{i}=
\langle 2^{+}_{i} \parallel \sum_{k}^{(p/n)} r^2_k
Y_2(\Omega_k)  \parallel g.s.\rangle.  
\end{equation}
If F-spin is preserved to a good extent, we must have
$R^{2}_{1}(IV/IS) <1$ and 
$R^{2}_{2}(IV/IS) >1$. 
\begin{table}
\caption{\label{tab1} Neutron and proton quadrupole transition amplitudes and 
IV/IS ratios for the lowest two $[2^{+}]_{RPA}$ states}
\begin{ruledtabular}
\begin{tabular}{cccccc}
$g.s. \rightarrow  [2^{+}_{i}]_{RPA}$ &$G_{2}$  &     ${\cal M}_{i}^{(n)}(fm^{2})$&
  ${\cal M}_{i}^{(p)}(fm^{2})$  &  $R_{i}(IV/IS)$& 
 $ R^{2}_{i}(IV/IS)$  \\ \hline

$g.s. \rightarrow  [2^{+}_{1}]_{RPA}$  &0.0 & 67.67
 &   41.98   & 0.234 & 0.055  \\
 &  1.0& 72.2
 &   49.6   &  0.185 & 0.034  \\\hline
 $g.s. \rightarrow  [2^{+}_{2}]_{RPA}$ & 0.0& 52.67
 &   46.40   &  0.063& 0.004  \\
 &1.0 & 14.4
 &   28.4   & - 0.327& 0.107 \\ 

\end{tabular}
\end{ruledtabular}
\end{table}
 
In order to meet all the above requirements we have only one 
parameter at our disposal, the quadrupole pairing constant.
This, while affecting little the lowest $[2^{+}_{1}]_{RPA}$, has an 
appreciable impact on the second $[2^{+}_{2}]_{RPA}$.
For $G_{2} = 0$, the ratios shown in Table \ref{tab1} 
qualify not only the first but also the
second $[2^{+}]_{RPA}$ as p-n symmetric. The second is, actually,  
even more symmetric and collective than the first. Moreover,
the corresponding one-phonon QPM state is pushed above 
a two-phonon state and becomes the third excited state, contrary to
the experimental data.  
As we increase $G_{2}$, the transition amplitudes, specially the 
neutron one, decrease,  but not enough.
Even for $G_{2}/\kappa = 1$, in fact, the neutron amplitude
is small but positive, so that the corresponding $IV/IS$ ratio
is much larger than in the case of vanishing $G_{2}$, but, still, 
appreciably smaller than one. Only for 
values of $G_{2}$ considerably larger than $\kappa$,
the $[2^{+}_{2}]_{RPA}$ becomes p-n non symmetric ($R(IV/IS) > 1$),
but looses completely its collectivity. The corresponding E2 strength 
is negligible, at variance with experiments. 
We therefore chose $G_{2} = \kappa$ which allows to  
fulfils  more closely the experimental requirements.

For such a  value, the lowest RPA
$2^{+}_{1}$ is collective, though to a less extent than in other
nuclei of the same region. 
Its RPA E2 decay strength (Table \ref{tab2}) is smaller
than the corresponding one in $^{94}$Mo \cite{LoSto02} by more 
than a factor two.
Such a quenching reflects the diminished role of the proton with
respect to the neutron component in $^{92}$Zr.  The neutron
dominance, however, is far less pronounced than in SM and does not
alter dramatically the symmetry of the state. In fact, not only all
proton and neutron components are in phase, but also the ratio of
the isovector to the isoscalar quadrupole transition amplitudes
is  small, though not negligible. Such a  test qualifies the
$2^{+}_{1}$  as a $\Delta T = 0$ p-n symmetric state with a small, 
though non negligible,
admixture of non symmetric pieces. 
\begin{table}
\caption{\label{tab2} Structure of the lowest RPA phonons (only the
largest components are given) and corresponding E2 reduced transition 
strengths in $^{92}$Zr. The states are normalized according to Eq. 
(\ref{phnorm})}
\begin{ruledtabular}
\begin{tabular}{ccccc}
$\lambda_i^\pi $ & $\omega_{\lambda_{i}^\pi} $(MeV)  &
$B(E2)\downarrow (w.u.)$ & Structure &\\
 \hline 2$_{1}^{+}$ & 1.18 &7.6&$0.99(2d_{5/2}\otimes 2d_{5/2})_n$ &48.4\%  \\
&&&$+0.23(2d_{5/2} \otimes 3s_{1/2})_n$  &4.8\%  \\
&&&$ +0.17(2d_{5/2} \otimes 1g_{9/2})_n$ & 2.5\%  \\
&&&total&60\%\\\hline
&&& $+0.64 (1g_{9/2} \otimes 1g_{9/2})_p$ &20.5\%   \\
&&&$ +0.23(1f_{5/2} \otimes 2p_{1/2})_p$  &5.5\%  \\
&&&$ +0.23(2p_{3/2} \otimes 2p_{1/2})_p$  &5.2\%  \\
&&&$ +0.15(1f_{5/2} \otimes 1f_{5/2})_p$ & 1.1\%  \\
&&&$ +0.14(1g_{9/2} \otimes 2d_{5/2})_p$  &2.1\%  \\
&&&total&40\%\\
 \hline 2$_{2}^{+}$ & 2.07&2.2&$-0.99(2d_{5/2}\otimes 2d_{5/2})_n$ &48.8\%  \\
&&&$+0.09(2d_{5/2} \otimes 3s_{1/2})_n$  &0.72\%  \\
&&&$+0.11 (2d_{5/2} \otimes 1g_{9/2})_n$ & 0.82\%  \\
&&&total&51\%\\\hline
&&& $ +0.79(1g_{9/2} \otimes 1g_{9/2})_p$ &30.7\%   \\
&&&$+0.25 (1f_{5/2} \otimes 2p_{1/2})_p$  &5.7\%  \\
&&&$+0.25 (2p_{3/2} \otimes 2p_{1/2})_p$  &5.7\%  \\
&&&$ +0.18(1f_{5/2} \otimes 1f_{5/2})_p$ & 1.3\%  \\
&&&$ +0.15(1g_{9/2} \otimes 2d_{5/2})_p$  &1.9\%  \\
&&&total&49\%\\
\end{tabular}
\end{ruledtabular}
\end{table}

The F-spin breaking is more substantial in the second $[2^{+}_{2}]_{RPA}$.
Indeed, its isoscalar
to isovector ratio $R_{2}(IV/IS)$  
is considerably smaller than unity, 
indicating that the transition is promoted with comparable strengths
by both isoscalar and 
isovector  quadrupole operators. 
This conclusion seems to be contradicted by the structure of the
wave function (Table \ref{tab2}), whose largest 
proton and neutron components  are in opposition of phase.
 
In order to solve this puzzle, we have analyzed more closely the 
neutron contributions to the $IV/IS$ ratio.
As shown in Table \ref{tab3}, the contribution of the 
largest neutron component to the matrix element of the quadrupole 
field, ${\cal M}^{(n)}_{2}$, is negative   
and opposite to the proton contribution. This neutron component would, 
therefore yield $R_{2}(IV/IS) > 1$, as expected.  
On the other hand, the remaining 
neutron configurations  are in phase with 
the protons (Table \ref{tab2}) and, 
because of the collective nature of the $2^{+}_{2}$, are in large number. 
Moreover, for most of 
these configurations, the two-quasiparticle matrix elements of the 
quadrupole field are quite sizeable (Table \ref{tab3})
and compensate for their small amplitude coefficients 
in the wave function. Their 
integrated contribution overcomes the one 
due to the dominant component, yielding a total $IV/IS$ ratio
smaller than one.  
\begin{table}
\caption{\label{tab3}   
Contribution of selected neutron configurations of the second RPA 
phonon $[2^{+}_{2}]_{RPA}$ to the matrix element ${\cal M}^{(n)}_{2} 
(RPA)$ of the quadrupole field in $^{92}$Zr. 
${\cal M}^{(n)}_{2} (q_{1}q_{2})$ gives 
the pure two-quasiparticle matrix elements.}
\begin{ruledtabular}
\begin{tabular}{cccccc}
$q_{1}\otimes q_{2} $ &${\cal M}^{(n)}_{2} (q_{1}q_{2})\, (fm^{2})$ 
&$\psi$& $\phi$&\%&${\cal M}^{(n)}_{2} (RPA)\, (fm^{2})$ \\
 \hline 
$2d_{5/2} \otimes 2d_{5/2}$  & 17.86 & -0.99& 0.09&48.8\%&-16.07 \\
$2d_{5/2} \otimes 3s_{1/2}$  & 25.33 & 0.10& 0.04&0.72\%&3.51 \\
$1g_{9/2} \otimes 2d_{5/2}$  & 34.85 & 0.11& 0.07&0.82\%&6.19 \\
$2p_{3/2} \otimes 2f_{7/2}$  & 33.17 & 0.02& 0.01&0.01\%&1.12 \\
$1f_{7/2} \otimes 1h_{11/2}$  & 57.68 & 0.05& 0.04&0.07\%&4.79 \\
$1g_{9/2} \otimes 1i_{13/2}$  & 72.83 & 0.05& 0.04&0.09\%&7.01 \\
$1d_{3/2} \otimes 1g_{7/2}$  & 33.83 & 0.02& 0.02&0.01\%&1.22 \\
  $\cdots$&$\cdots$&$\cdots$&$\cdots$&$\cdots$&$\cdots$\\
\end{tabular}
\end{ruledtabular}
\end{table}
The breaking of the p-n symmetry is
therefore the result of a competition between a dominant 
configuration, with relatively small quadrupole matrix 
elements, and a large number of small components yielding 
large quadrupole transition amplitudes and acting coherently.  
We may, therefore, infer from this RPA analysis that F-spin is 
more severely broken in the second rather than the lowest
$2^{+}$. Our findings are quite different from the SM results.
These discrepancies can
be explained with the severely truncated space used in the SM,
which is spanned only by two neutrons and two protons external to
the inert cores $N_c =50$ and $Z_c = 38$, respectively. As shown in tables
\ref{tab2} and \ref{tab3}, this truncation excludes  
many configurations whose contribution change dramatically 
the nature of the second $[2^+_{2}]_{RPA}$ state.

\subsection{QPM results}

Let us now investigate the QPM states
(Table \ref{tab4}) and how their phonon composition affects
the E2 (Tables \ref{tab5}) as well as the E1 and M1 transitions 
(Table \ref{tab6}). 
\begin{table}
\caption{\label{tab4} Energy and phonon structure of selected
low-lying excited states in $^{92}$Zr. Only the dominant
components are shown.}
\begin{ruledtabular}
\begin{tabular}{ccccc}
\multicolumn{2}{c} {State} &
\multicolumn{2}{c}{E (keV)} & Structure,\%  \\
  &  J $^\pi$ &  EXP &  QPM  &     \\ \hline

& 2$_{1}^{+}$  &  934  &  1017 & $92.6\%[2_{1}^{+}]_{RPA}$  \\

& 2$_{2}^{+}$ &  1847  &  1945  &
$90\%[2_{2}^{+}]_{RPA}$ \\

& 2$_{3}^{+}$ &  2067  &  2008  &
$19\%[2_{3}^{+}]_{RPA} + 65\%[2_{1}^{+} \otimes 2_{1}^{+}]_{RPA}$ \\

& 2$_{7}^{+}$ &     &  3386  &
$32\%[2_{5}^{+}]_{RPA} + 51\%[2_{1}^{+} \otimes 2_{2}^{+}]_{RPA}$ \\

& 4$_{1}^{+}$ &  1495  & 1598  & $78\%[4_{1}^{+}]_{RPA}+
17\%[2_{1}^{+} \otimes
2_{1}^{+}]_{RPA}$  \\

& 4$_{2}^{+}$ &  2398  & 2188  & $18\%[4_{1}^{+}]_{RPA} +
59\%[2_{1}^{+} \otimes
2_{1}^{+}]_{RPA}$  \\

& 4$_{3}^{+}$ &  2863  & 2906  & $27\%[4_{2}^{+}]_{RPA}
+32\%[2_{1}^{+} \otimes 4_{1}^{+}]_{RPA}+ 28\%[2_{1}^{+} \otimes
2_{2}^{+}]_{RPA}$
\\\hline

& 1$_{1}^{+}$ &  3472  & 3235 &
$ 93\%[2_{1}^{+} \otimes 2_{2}^{+}]_{RPA}+6\%[1_{1}^{+}]_{RPA}$   \\

& 1$_{2}^{+}$ &  3638  & 3781 &
$ 91\%[1_{1}^{+}]_{RPA}$   \\

& 3$_{2}^{+}$ &   & 3180  & $74\%[2_{1}^{+} \otimes
2_{2}^{+}]_{RPA}$  \\\hline

& 1$_{1}^{-}$ &  3370  & 3398 &
$ 99.3\%[2_{1}^{+} \otimes 3_{1}^{+-}]_{RPA}$   \\

& 3$_{1}^{-}$ &2339   & 2387  & $81\%[3_{1}^{-}]_{RPA}+
14\%[2_{1}^{+} \otimes
3_{1}^{-}]_{RPA}$  \\
 \end{tabular}
\end{ruledtabular}
\end{table}

The first $2^{+}_{1}$ is  mostly accounted for by the
lowest RPA one-phonon component. The appreciable, but not overwhelming,
neutron dominance is therefore confirmed  and is
consistent with the magnetic properties of the state. Indeed, the QPM yields
for the gyromagnetic factor $g(2_1^+) =-0.20$, very close to the
experimental value $g_{exp}(2_1^+) =-0.18(1)$. Apparently, 
the experimental negative g-factor claims only a moderately 
large neutron weight.

The second $2^{+}_{2}$ is a one-phonon state, dominated by the second 
$[2^{+}_{2}]_{RPA}$. As pointed out already, this is peculiar of $^{92}$Zr,
since the $2^{+}_{2}$  in the nearby nuclei $^{94}$Mo and
$^{136}$Ba was found to be a
two-phonon symmetric state \cite{LoSto02}. 
This one-phonon $2^{+}_{2}$ undergoes 
an E2 decay to the ground state (Table \ref{tab5}) and a M1 transition 
to the symmetric $2^{+}_{1}$ (Table \ref{tab6}) . 
The computed M1 strength, though fairly large, is smaller
than in $^{94}$Mo, while the E2 strength is unusually large. These
two correlated facts, already pointed out in the experimental
analysis \cite{Werner02}, indicate that the $2^{+}_{2}$ is not a
pure "mixed symmetry" state but has an appreciable F-spin
symmetric component. The latter piece, in fact, is responsible for
the enhancement of the E2 and quenching of the M1 strengths,
respectively, with respect to $^{94}$Mo. The breaking of F-spin is
even more pronounced than what predicted by the present QPM
calculation. This, in fact, overestimates the experimental M1
strength and underestimates the E2 transition probability roughly
by the same factor $\sim 1.5$. 
Our QPM result is therefore at variance with the SM 
findings. This discrepancy emerges clearly from the analysis of the 
magnetic moments. The QPM g-factor for the
$2^{+}_{2}$ state is $g(2_2^+) = -0.31$, quite different in sign
and magnitude from the SM value $g_{SM}(2_2^+) = 0.9$. Clearly, a
measure of this quantity would discriminate between the two
descriptions and, therefore, would shed light on the p-n symmetry
of this state.
\begin{table}
\caption{ \label{tab5} $E2$ transitions connecting some excited
states in $^{92}$Zr calculated in QPM. The experimental data are
taken form Ref. \protect\cite{Werner02}}
\begin{ruledtabular}
\begin{tabular}{ccc}
      $B(E2; J_{i} \rightarrow J_{f})(w.u.)$&
  EXP &  QPM  \\ \hline

 $  B(E2; 2^{+}_{1} \rightarrow g.s.)$
 &   6.4(6)   &  6.5  \\

  $  B(E2; 2^{+}_{2} \rightarrow g.s.)$
 &   3.7(8)   &  2.1  \\

 $ B(E2;2^{+}_{2} \rightarrow  2^{+}_{1})$
   &  0.3(1) & 0.39 \\

$ B(E2;2^{+}_{3} \rightarrow  2^{+}_{1})$
   &    & 6.8 \\

 $ B(E2; 4^{+}_{1} \rightarrow 2^{+}_{1}) $
   &  4.04(12) &  1.0 \\
 $ B(E2; 4^{+}_{2} \rightarrow 2^{+}_{1}) $
   &    &   8.4 \\
   $ B(E2; 4^{+}_{7} \rightarrow 2^{+}_{1}) $
   &    &  2.4\\
   $ B(E2; 4^{+}_{7} \rightarrow 2^{+}_{2}) $
   &    &  2.6\\\hline

 $ B(E2; 1^{+}_{1} \rightarrow 2^{+}_{1})$
 & $<$8.0(6)  &   2.0  \\

 $ B(E2; 3^{+}_{2} \rightarrow 2^{+}_{1}) $
 &  &   1.6  \\
 $ B(E2; 3^{+}_{2} \rightarrow 2^{+}_{2}) $
 &  &   4.4 \\
\end{tabular}
\end{ruledtabular}
\end{table}

Another  distinguishing feature of $^{92}$Zr with respect to the
nearby nuclei is the fragmentation of the QPM states into several
multiphonon components. The symmetric two-phonon $\left[ 2^+_{1}
\otimes 2^+_{1} \right] _{RPA}$ accounts only for 65\% of the
$2^{+}_{3}$. For the sake of comparison, the
two-phonon counterpart in $^{94}$Mo represents the 82\% of the
$2^{+}_{2}$ state. The  non symmetric $\left[ 2^+_{1} \otimes
2^+_{2} \right] _{RPA}$ is spread over several $2^{+}$ states. It
is dominant in the $2^{+}_{7}$ and sizeable in $2^{+}_{8}$. These
two states are therefore predicted to have strong E2 decays to the
p-n non symmetric $2^{+}_{2}$ state. The few available data are
closely reproduced by the calculation. Unfortunately, most of the
predicted strong transitions have not been detected yet.

The $\left[
2^+_{1} \otimes 2^+_{1} \right] _{RPA}$ is spread among several
$4^{+}$ QPM states, thereby generating several $4^{+}$ excitations
with appreciable, if not strong,
E2  decay. The experiments have detected only one $4^{+}$ and
measured the strength of its E2 transition to the $2^{+}_{1}$ as
well as its g-factor, which resulted to be $g_{exp}(4^+) =-0.50$.
It is unlikely that this excitation corresponds to the first QPM 
$4^{+}_{1}$  dominated by the RPA
one-phonon $\left[4^{+}_{1}\right]_{RPA}$  with an appreciable
admixture (17\%) of  the two-phonon component $\left[ 2^+_{1}
\otimes 2^+_{1} \right]_{RPA}$. Its E2 decay strength  to the
$2^{+}_{1}$ is a factor four smaller than the measured value and
its g-factor is $g(4_1^+) =0.57$, opposite in sign to the
experimental value. It is more natural to associate the observed 
excitation to the second $4^{+}_{2}$,
composed of a dominant two-phonon component with a small one-phonon
admixture.  Its E2 strength is only a factor two larger than
the measured value and its g-factor is $g(4_2^+) =-0.32$,
reasonably close to the experimental value. This comparative analysis  
claims also a more pronounced fragmentation leading to
a further quenching of the E2 decay strength of the  $4^{+}_{2}$ and 
to an enhancement of its g-factor. On
the other hand,  a further fragmentation would enhance the amplitude
of the two-phonon component $\left[ 2^+_{1} \otimes 2^+_{1}
\right]_{RPA}$ and, therefore, the E2 decay strength of the first
$4^{+}_{1}$. The observation of additional $4^{+}$ states and the
measurement of their  E2 decays would represent a precious test
for testing the multiphonon fragmentation of the states predicted
in QPM.

The $1^{+}$ and $3^{+}$ states are also of importance
for testing the p-n symmetry. As shown in Table \ref{tab4}, only
the first $1^{+}_{1}$ is predominantly a two-phonon non symmetric
state and, therefore, would be the analogue of the IBM
mixed-symmetry state, if F-spin were conserved. The other  has
a dominant spin excitation component. Spin indeed contributes mainly to
the strength of the $M1$ decay of the second $1^{+}_{2}$. It gives 
also a small but non negligible contribution to the decay of the  $1^{+}_{1}$.  
Such a contribution is crucial for attaining a good agreement 
with experiments.  
The strong E2 decay of the $1^{+}_{1}$
to the symmetric $2^{+}_{1}$ is also consistent with the
experiments and mirrors the unusually strong E2 decay of the
$2^{+}_{2}$ to the ground state. It represents, therefore, an
additional signature of F-spin breaking. A further confirm  
may be provided by the E2 decay of the $3^{+}_{2}$. This contains a very large
$\left[ 2^+_{1} \otimes 2^+_{2} \right] _{RPA}$ component and
is predicted to decay to the $2^{+}_{1}$ with a strong E2 transition. An
experimental test would be desirable.

Very intriguing is the case of the $2_{3.263}^{+}$ level observed
at $E=3.263$ MeV. This state decays to the first $2_1^+$ with an
appreciable M1 strength and an E2 strength of the order of one
single particle unit. This level is 
close to the energy $E\simeq 3.7$ MeV of the three phonon state  
$ \mid 2_{3ph}^+ \rangle = 
\left[2^+_{1} \otimes 2^+_{1} \otimes 2^+_{2} \right]_{2^+}$. 
Moreover, the strength of the  
M1 transition of this three phonon state 
to the first $2_1^+$ has
the same structure of and is comparable in magnitude to the
strength of the M1 decay of the  non
symmetric $1^{+}_{1}$ to the ground state. For a pure, properly 
antisymmetrized, three-phonon state, we get  
$B(M1,2_{3ph}^+ \rightarrow 2_1^+) = 0.06 \mu_N^2 $,  close to
$B(M1,1_{1}^+ \rightarrow 0_{gr}) = 0.07 \mu_N^2 $ and smaller 
than the measured strength by a factor two. Thus, it is  
tempting to consider this $2_{3.263}^{+}$  as a good candidate
for being a three-phonon excitation with small admixture of two-phonon
components. If confirmed by more complete calculations, this level would 
provide the first evidence of a
three-phonon non symmetric $2^+$ state.
\begin{table}
\caption{\label{tab6} QPM versus experimental $M1$, $E1$, and $E3$ transitions
between some excited states in $^{92}$Zr. The experimental data
are taken from Ref. \protect\cite{Werner02} }
\begin{ruledtabular}
\begin{tabular}{ccccc}
   &$ B(M1; J_{i} \rightarrow J_{f}) (\mu_N^{2}) $  & EXP & QPM& 
   QPM ($g_{s}=0$) \\\hline

& $B(M1; 2^{+}_{2} \rightarrow 2^{+}_{1})$  &
 0.46(15)  &  0.68& 0.22   \\

 & $B(M1; 2^{+}_{3.263} \rightarrow 2^{+}_{1})$  &
 0.16(2)  &  0.06\footnotemark[1]  \\

 &$  B(M1; 1^{+}_{1} \rightarrow   g.s.)$   &
 0.094(4) &  0.069& 0.031  \\

 &$  B(M1; 1^{+}_{2} \rightarrow  g.s.)$ &  &
 0.081 &  0.018 \\ 
&$  B(M1; 1^{+}_{1} \rightarrow  2^{+}_{1})$   &
  $<0.089(6)$ &  1 $\times  10^{-4}$ &  5 $\times  10^{-5}$\\\hline
 &$ B(E1; J_{i} \rightarrow J_{f})  $  &   &  & \\\hline
& $B(E1; 1^{-}_{1} \rightarrow g.s.)$  &
 $0.037(4)\times10^{-3}(e^{2}fm^{2}$)  & $0.045\times10^{-3}  
 (e^{2}fm^{2}$) & \\
& $B(E1; 3^{-}_{1} \rightarrow 2^{+}_{1})$  &
   & $ 0.43\times10^{-3}$  ( w.u.)  &\\
   & $B(E1; 3^{-}_{1} \rightarrow 2^{+}_{2})$  &
   & $ 1.9\times10^{-3}$  ( w.u.)  &\\
   & $B(E3; 3^{-}_{1} \rightarrow  g.s.)$  &
   &   24  ( w.u.)  &\\
\end{tabular}
\end{ruledtabular}
\footnotetext[1]{Computed under the assumption that the state is a
pure, properly antisymmetrized, three-phonon state.}
\end{table} 

 Valuable pieces of information come from the study of the low-lying
 negative parity states. Consistently with previous QPM calculations
 \cite{PonSto}, the first $1^{-}_{1}$ is a pure
  $\left[ 2^+_{1} \otimes 3^-_{1} \right] _{RPA}$ two-phonon state.
 The computed energy is higher than the measured value and close to
 the sum of the $2^{+}_{1}$ and $3^{-}_{1}$ energies. The strength of
 the E1 decay to the ground state is in good agreement with  
 experiments and results from a destructive interference between the surface
 quadrupole-octupole mode and the IVGDR.
 The $3^{-}_{1}$ is predominantly an octupole mode and is predicted to
 undergo a strong E3 decay to the ground state.

 Because of the isovector nature of the electric dipole operator, the
 E1 transitions of the $3^-_{1}$ to the two lowest $2^{+}$ states  may
 be used as a test of their p-n symmetry. The ratio $R_{E1} 
 =B(E1;3_{1}^{-}\rightarrow 2_{2}^{+})/B(E1;3_{1}^{-}\rightarrow 
 2_{1}^{+})$
 was measured recently \cite{Pietr03} and found to be $R_{E1} =2.7(2)$,
 larger than 1 but an order of magnitude smaller than in the nearby 
 nuclei like $^{94}$Mo.  The authors 
 of Ref. \cite{Pietr03}, driven by the truncated SM calculation,  
 ascribed this suppression to the neutron 
 character of the first $2_{1}^{+}$.
 Our QPM calculation yields for this ratio $R_{E1}= 4.4$, larger than 
 the experimental value by only a factor $\sim 1.5$. This rather 
 satisfactory agreement suggests that the suppression might be due 
 to F-spin breaking in  both $2^{+}_{1}$ and, specially,  $2^{+}_{2}$ 
 states. Moreover,
 the overestimation of the experimental ratio suggests a more pronounced 
 p-n symmetry breaking than in our QPM scheme.   

\section{Concluding remarks}

On the ground of the present study, we may draw the conclusion
that, consistently with the experimental analysis \cite{Werner02},
the lowest two $2^{+}$ are RPA one-phonon states. At variance with
the conclusion drawn in \cite{Werner02}, based
on a calculation carried out within a too severely truncated SM
space, we found that the $2^{+}_{1}$ state  has appreciable but
not huge neutron dominance which does not destroy its
p-n symmetric character. The F-spin, instead, is broken more 
substantially in the second $2^+$ state.  A selective
comparison of the QPM with the experimental transition strengths
strongly suggests that this breaking is even more pronounced than what
predicted by the QPM calculation. Such a F-spin admixture provides
the key for a consistent description of all experimental levels
and transitions. The measure of selected additional transitions, as 
pointed out in the text, would
greatly contribute to a conclusive clarification of this issue.

Another remarkable result is the fragmentation of QPM states among
several multiphonon components. Because of such a spread, the
corresponding IBM states should also be linear combinations of
several multi-bosonic basis states.

Last but not least, the present study offers  arguments in
favor of the first experimental evidence of a three-phonon non
symmetric $2^+$ state.

\section{acknowledgements}
The present work was partly supported by the Italian Ministry of
Instruction, University, and Research (MIUR) and by
the Bulgarian Science Foundation (contract n. 1311). The authors
thank N.  Pietralla  and U. Kneissl for useful information and
discussions.

\end{document}